\documentclass[10pt,nofootinbib,amsmath,amssymb,showpacs,twocolumn]{revtex4}
\usepackage[T1]{fontenc} \usepackage[latin1]{inputenc}
\usepackage{amsmath,color} 
\usepackage{graphicx}
\usepackage{epsfig} \usepackage{amssymb} 
\usepackage[capitalize]{cleveref}

\begin{document}

\title{Structural transitions of skyrmion lattices in synthetic antiferromagnets}
\author{E. van Walsem$^1$}
\author{R. A. Duine$^{1,2}$}
\author{J. Lucassen$^2$}
\author{R. Lavrijsen$^2$}
\author{H.J.M. Swagten$^2$}
\affiliation{$^1$Institute for Theoretical Physics, Universiteit Utrecht, Leuvenlaan 4, 3584 CE Utrecht, The Netherlands}
\affiliation{$^2$ Department of Applied Physics, Eindhoven University of Technology, P.O. Box 513, 5600 MB Eindhoven, The Netherlands}

\begin{abstract}
Thin magnetic films with Dzyaloshinskii-Moriya interactions are known to host skyrmion crystals, which typically have a hexagonal lattice structure. We investigate skyrmion-lattice configurations in synthetic antiferromagnets, i.e., a bilayer of thin magnetic films that is coupled antiferromagnetically. By means of Monte-Carlo simulations, we find that by tuning the interlayer coupling the skyrmion lattice structure can be tuned from square to hexagonal. We give a simple interpretation for the existence of this transition based on the fact that for synthetic antiferromagnetic coupling the skyrmions in different layers repel each other and form each others' dual lattice. Our findings may be useful to experimentally switch between two lattice configurations to, for example, modify spin-wave propagation.
\end{abstract}

\pacs{PACS NRS}

\maketitle

\section{Introduction \label{secIntro}}



Skyrmions in magnetic materials have been attracting great interest recently. Skyrmions were first introduced in particle physics and  correspond to a classical stationary solution of the equations of motion with which a topological invariant is associated \cite{Skyrme1962}. Later, such topological configurations were considered in magnetic systems by Bogdanov and Hubert \cite{Bogdanov1994}. Skyrmions can be small, in the nanometer range, and behave as (pseudo)particles that can be moved without decaying. M\"uhlbauer \emph{et al}. demonstrated their existence with neutron scattering in MnSi in 2009 \cite{Muhlbauer2009}. Later, thin magnetic multilayers proved to be able to possess magnetic skyrmions \cite{Yu2010,Heinze2011,Dupe2014}. Initially, skyrmions were found at low temperatures but recently also room-temperature skyrmions have been created experimentally \cite{Yu2016,Woo2016,Soumyanarayanan2017,Moreau-Luchaire2016,Chen2015}. Due to their solitonic behaviour and their rigidity originating from their topological properties they hold great promise for information technologies \cite{Parkin2008,Kiselev2011,Fert2013}. For example, skyrmion race track memory systems are being developed in which skyrmions act as bits and in which their postion is manipulated by current \cite{Fert2013,Loth2012,Jonietz2010}. 

One of the classes of materials for hosting skyrmions are thin magnetic multilayers, such as Co/Pt, with high perpendicular magnetic anisotropy (PMA). The PMA is stronger than the in-plane anisotropy thus providing opportunities for skyrmions since the spin in the centre of a skyrmion is oriented perpendicular to the layer as well. Furthermore, Dzyaloshinskii-Moriya interactions (DM interaction or DMI) are essential for forming skyrmions in thin magnetic films \cite{Dzyaloshinsky1958,Moriya1960}. DMI needs two conditions to form: breaking of inversion symmetry and spin-orbit coupling. The stacking of different materials in the multilayers satisfies the first condition. To comply with the second condition, heavy metals such as Pt or Ta are used in the multilayers \cite{Fert1991}. DMI can have different forms. The interfacial DMI that arises in magnetic multilayers stabilises N\'eel skyrmions, whereas the bulk DMI, in e.g. MnSi, stabilises Bloch skyrmions. 

Examples of multilayer systems are synthetic antiferromagnets (SAFs) \cite{Duine2017}. SAFs consist of two ferromagnetic multilayers which are coupled antiferromagnetically through the spin-dependent RKKY coupling \cite{Salamon1986,Majkrzak1986,Grunberg1986}. By changing the thickness of the spacer, the coupling between the layers is oscillatory from antiferromagnetic to ferromagnetic and the magnitude is also dependent on the thickness of the spacer \cite{Parkin1990,Parkin1991,Bruno1995,Slonczewski1989,Faure-Vincent2002,Edwards1991}. Additionally, it has been shown that the coupling can be tuned with an external electric field \cite{Fechner2012}. The interlayer exchange coupling is much weaker than the exchange coupling within the layers so the antiferromagnetic order can compete with external fields and anisotropy. Zhang \emph{et al.} have shown that SAFs are promising for developing the skyrmion race track memory because the Magnus force, which influences the direction of a moving skyrmion, is opposite in the different layers and thus cancels out \cite{Zhang2016}. Furthermore, exchange coupling between layers has been shown to stabilise skyrmions \cite{Nandy2016,Chen2015}. 

In this paper we consider skyrmion lattices in synthetic antiferromagnets. We find that skyrmion lattices still occur for both synthetic ferromagnetic and synthetic antiferromagnetic coupling, although in the synthetic antiferromagnetic case a bigger magnetic field is needed to stabilise skyrmions. In addition, we find a structural phase transition in the synthetic-antiferromagnetic case from a triangular lattice to a square lattice. Skyrmion lattices in multilayer systems usually have a triangular configuration \cite{Bogdanov1994,Muhlbauer2009,Yu2010} although examples of square lattices also exist \cite{Heinze2011}. 
Similar transitions between triangular and square skyrmion lattices under magnetic field have been reported before in MnSi \cite{Nakajima2017} and in 2D layers \cite{Lin2015}. In $\beta$-Mn-type Co$_8$ Zn$_8$ Mn$_4$ the same transition is found for cooling in an applied magnetic field \cite{Karube2016}.
Analogous to skyrmions, similar transitions between triangular and square lattices are observed in vortex lattices in two component Bose Einstein Condensates, where the two components play the role of the two layers \cite{Mueller2002}. 
In contrast to the examples above, the structural phase transition in SAFs has a possibility to be tuned \emph{in situ} by altering the coupling between the layer by an electric field. Furthermore, in systems where also the DMI varies between the layers more exotic lattice configurations are achievable. Our research thus forms a route to tailoring skyrmion-lattices configurations. 

In  \cref{secMethod} we discuss our model and the simulations. After this we show results of our simulations in  \cref{secResults} and in  \cref{secDisc} we discuss these results. Finally, we conclude with an outlook in \cref{secConcl}.


\section{Model and method \label{secMethod}}

We model the two magnetic layers by classical Heisenberg spins with ferromagnetic nearest neighbour coupling $J_{\rm xy}$, easy-axis anisotropy $K$ along $z$ and the Dzyaloshinskii-Moriya interaction $D$. Furthermore, both layers experience a uniform magnetic field $B$ 
and are coupled to each other with interlayer exchange $J_{\rm z}$. In each layer the spins are positioned on the basis vectors of the simple square lattice ${\bf \hat{x}}$, ${\bf \hat{y}}$. We apply periodic boundary conditions in the $x$ and $y$-direction and open boundary conditions in the $z$-direction. The Hamiltonian is given by the sum of an interlayer and an intralayer part:
\begin{align}\label{eq:hamiltonian}
H = &\ H_{\rm intra} + H_{\rm inter} .
\end{align}
The intralayer part is expressed as:
\begin{align}\label{eq:hamiltonianINTRA}
H_{\rm intra} =
	-& J_{\rm xy} \sum_{\alpha \in \{T,B\}} \sum_{\bf r}  {\bf S}^{\alpha}_{\bf r}  \cdot \left(  {\bf S}^{\alpha}_{{\bf r}+{\bf \hat{x}}} +  {\bf S}^{\alpha}_{{\bf r}+{\bf \hat{y}}} \right) \nonumber \\ 
	+& K \sum_{\alpha \in \{T,B\}} \sum_{\bf r}  \left( {\bf S}^{\alpha}_{\bf r} \cdot {\bf \hat{z}}  \right)^2 - {\bf B} \cdot \sum_{\alpha \in \{T,B\}} \sum_{\bf r}  {\bf S}^{\alpha}_{\bf r} \nonumber  \\
	-& D \sum_{\alpha \in \{T,B\}} \sum_{\bf r} \left(  {\bf S}^{\alpha}_{\bf r} \times {\bf S}^{\alpha}_{{\bf r}+{\bf \hat{x}}} \cdot {\bf \hat{y}} - {\bf S}^{\alpha}_{\bf r} \times {\bf S}^{\alpha}_{{\bf r}+{\bf \hat{y}}} \cdot {\bf \hat{x}} \right),  
\end{align}
where ${\bf S}_{\bf r}^{\alpha}$ denotes a normalised spin at position $\bf{r}$ in either the top ($\alpha=T$) or bottom ($\alpha=B$) layer. For simplicity we assume ${\bf B} = B  {\bf \hat{z}}$ to be aligned with the $z$-axis. The interlayer part is given by:
\begin{align}\label{eq:hamiltonianINTRA}
	H_{\rm inter} = 	-& J_{\rm z} \sum_{\bf r}  {\bf S}^{T}_{\bf r}  \cdot  {\bf S}^{B}_{\bf r+ \hat{z}}. 
\end{align}

We use Monte Carlo simulations to investigate the ground state of this model. In our Monte Carlo simulation a random configuration of spins is generated. In one Monte Carlo step we pick a random spin and propose a new vector its direction. This vector is chosen from a cap which size is chosen such that the acceptance rate in our Metropolis algorithm is 50\%. 
To reach the ground state and avoid getting trapped in local minima, we use simulated annealing in which we thermalise a system at a high temperature and subsequently lower the temperature stepwise until a temperature close to zero is reached. This proces is done multiple times. 


\section{Results \label{secResults}}
\subsection{Phases}
\begin{figure}
\includegraphics[width=\linewidth]{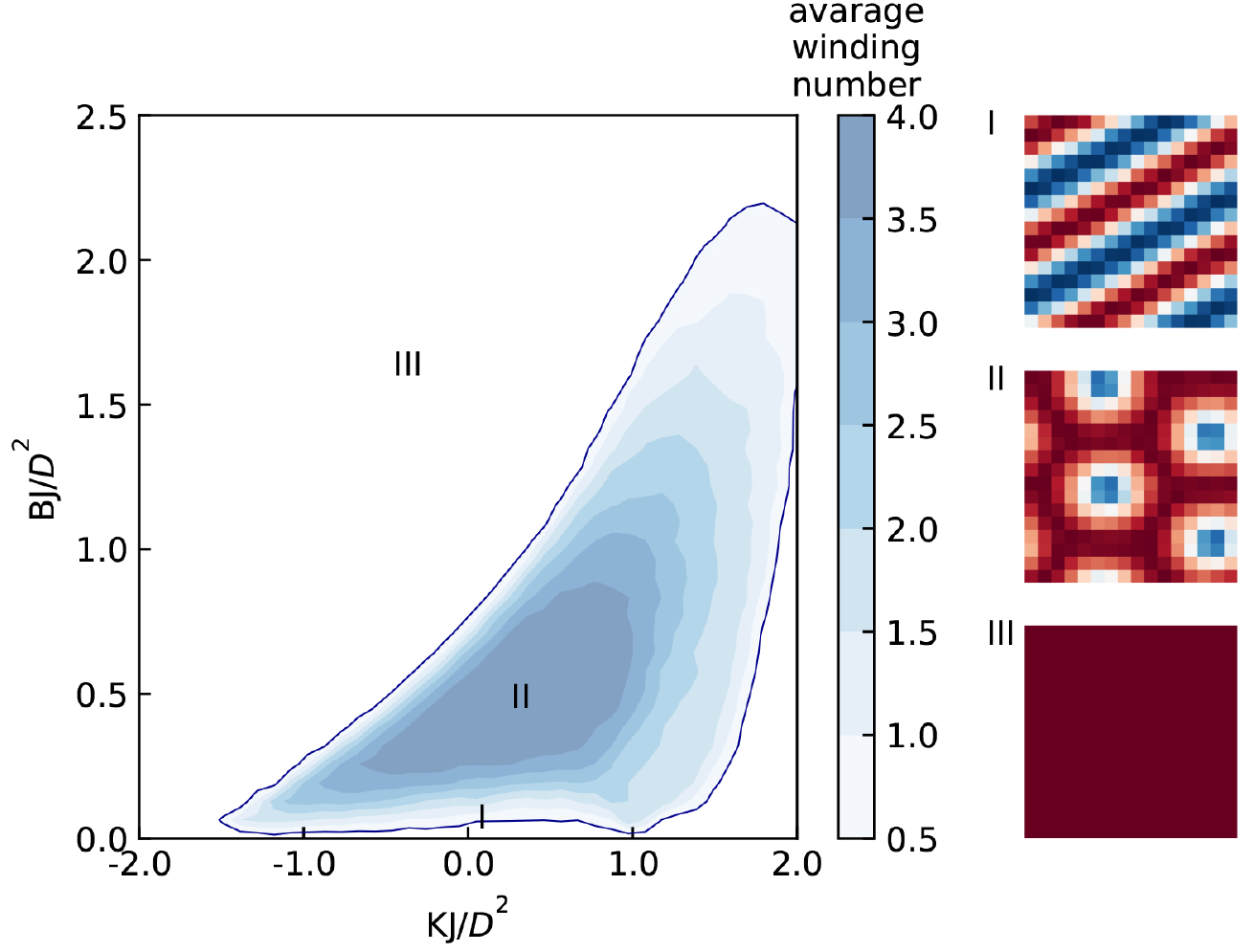}
\caption{Phase diagram of the skyrmion pocket for $BJ_{\rm xy}/D^2$ versus $KJ_{\rm xy}/D^2$ with $J_{\rm z}/J_{\rm xy}$=0. The colour scale depicts the winding number. The outer contour line is at winding number equals 0.5.  I, II, III depicts the phases found on the given locations in the phase diagram, snapshots and phase diagram are from a 16x16 system with a pitch of $p=8$. The colour in the snapshots depicts the $z$-component of the spin using the colorbar in \cref{figFFT}. 
} 
\label{figPhaseDiagram}

\end{figure}

In this section we address the three types of phases in our simulations: the spiral state, skyrmion state and fully polarised state. In the spiral state the DMI and exchange dominate. The direction of the spirals is determined by the DMI, and the wavelength is determined by the ratio of the DMI and intralayer exchange coupling. See \cref{figPhaseDiagram} I for a snapshot of a z-projection of a spiral state. 
For an increased external magnetic field we find the skyrmion state. 
Skyrmions form in lattices and between the skyrmions spins point along the external magnetic field, as shown in \cref{figPhaseDiagram} II. High external magnetic field can suppress the formation of skyrmions and subsequently systems containing a small number of skyrmions, without lattices structure, are found as well. The size of a skyrmion is determined by the ratio of the DMI and intralayer exchange coupling and is called the skyrmion pitch size $p$ which is given by $D/J_{\rm xy} = \tan(2\pi / p)$. 
In the polarised state the external magnetic field dominates over the anisotropy and DMI and all the spins in the system point in the same direction as the external magnetic field as shown in \cref{figPhaseDiagram} III. 

We determine the skyrmion state from the winding number. The winding number (also called chirality or topological charge) is an integer which represents the number of times the spins enclose an unit sphere. The skyrmion state is recognised as the system with a nonzero winding number. In the continuum limit the winding number is expressed as:
\begin{equation} 
w = \frac{1}{4\pi}\int dx\ dy\ {\bf m}\cdot \left(\frac{\partial {\bf m}}{\partial x}\times\frac{\partial {\bf m}}{\partial y} \right),
\end{equation}
where $x,y$ are the directions within the plane and 
${\bf m}=\langle {\bf S} \rangle / |\langle {\bf S} \rangle|$ is the magnetisation direction in the system. 
In \cref{figPhaseDiagram} we show the phase diagram, and typical spin configurations of the phases. Systems with a nonzero winding number possess skyrmions and thus are indicated as such in the phase diagram. In the phase diagram we plotted a contour plot of the total winding number of the system. In our systems the skyrmions have a winding number of positive one so the total winding number of a system is a measure for the total number of skyrmions in the system.

\begin{figure}
\includegraphics[width=\linewidth]{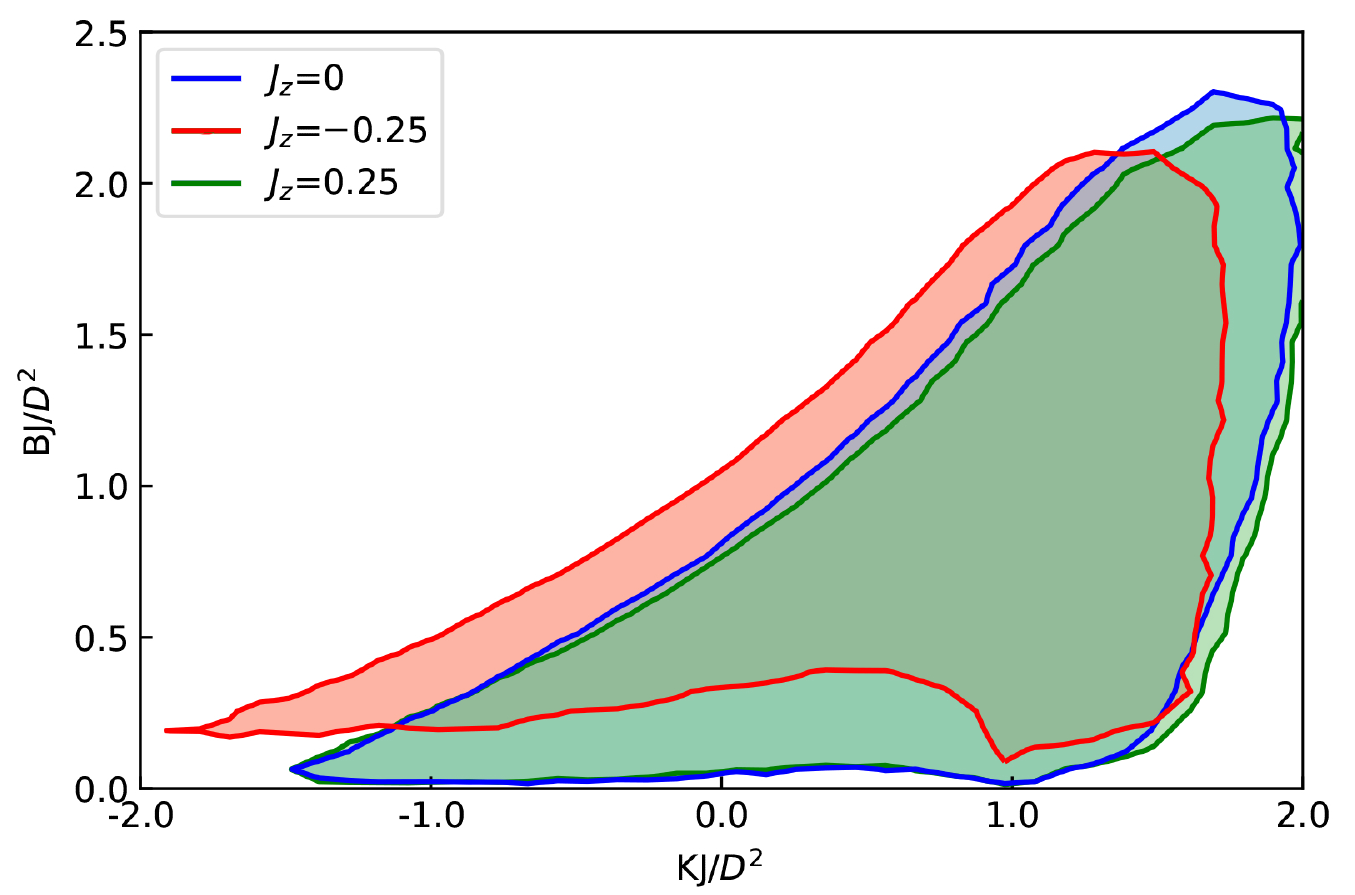}
\caption{Phase diagram of the skyrmion pocket for $BJ_{\rm xy}/D^2$ versus $KJ_{\rm xy}/D^2$ with $J_{\rm z}/J_{\rm xy} = 0$ (blue), $-0.25$ (red) and $0.25$ (green). The skyrmion pocket is drawn for systems with a total winding number $w>0.5$. It shows that the skyrmion pocket for  $J_{\rm z}/J_{\rm xy} = 0,$ and $0.25$ overlay, and that the skyrmion pocket for $J_{\rm z}/J_{\rm xy} = -0.25$ occurs for higher magnetic field. The simulated system had a system size of 16x16 spins with a pitch of $p=8$.
} 
\label{figInterlayerPhaseDiagram}
\end{figure}

Upon including the interlayer exchange, we see that in the synthetic ferromagnetic region there is no significant difference compared to the non-interacting picture as shown in figure \cref{figInterlayerPhaseDiagram}. Looking at the synthetic antiferromagnetic region we see that the skyrmion pocket is situated at higher external magnetic field. 
This is explained as follows: the spiral state can adjust to the synthetic antiferromagnetic coupling by shifting the spirals in different layers half a period with respect to each other without increasing the energy. On the contrary, the polarised phase cannot adjust in a comparable way and needs a higher external magnetic field to adjust to the synthetic antiferromagnetic coupling. The skyrmion state is partly polarised while also having a lattice which can shift in respect to each other. Our simulations show that the lower bound of the skyrmion pocket rises considering the energetic advantage of the spiral state. The higher bound of the pocket rises as well because of the energy advantage of the skyrmion lattices over the polarised state originating from the shifting of the skyrmion lattices with respect to each other. 


\subsection{Structural phase transition}
\begin{figure}
\includegraphics[width=0.9\linewidth]{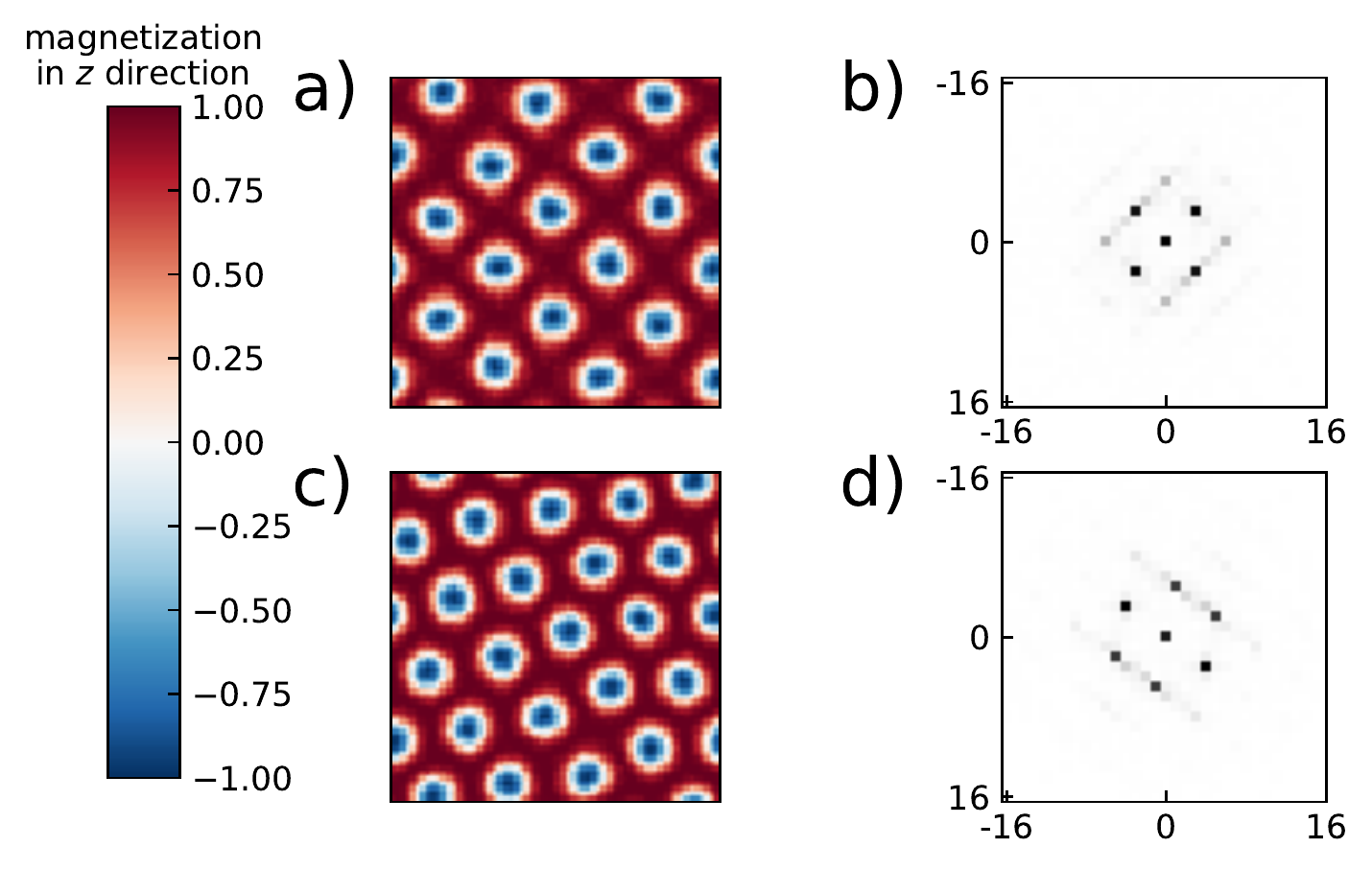}
\caption{Spin configurations of a 64x64, $p=13$, $B=0.56$, $K=0$, system showing {\bf a)} square lattice configuration with $J_{\rm z}=-0.14$, {\bf b)} the two dimensional Fourier transform of the square lattice configuration {\bf c)} hexagonal lattice configuration with $J_{\rm z}=0.01$. {\bf d)} the two dimensional Fourier transform of the square lattice configuration. From the Bragg peaks the lattice configuration is classified. }
\label{figFFT}
\end{figure}

Taking a closer look at the synthetic antiferromagnetic skyrmion phases we notice that the lattice configurations change from a triangular lattice, \cref{figFFT} c), to a square one \cref{figFFT} a).  
We determine the lattice configurations of the skyrmions by looking at the reciprocal lattice which we obtain by taking a two dimensional Fourier transform of the configuration. A square lattice shows four equally distributed first order peaks, as shown in \cref{figFFT} b), where a hexagonal lattice shows six \cref{figFFT} d). The spiral phase is recognisable with having a zero winding number and two peaks with an angle of 180 degrees between each other. The fully polarised state has no nonzero winding number nor a nontrivial reciprocal lattice.

\begin{figure}
\includegraphics[width=\linewidth]{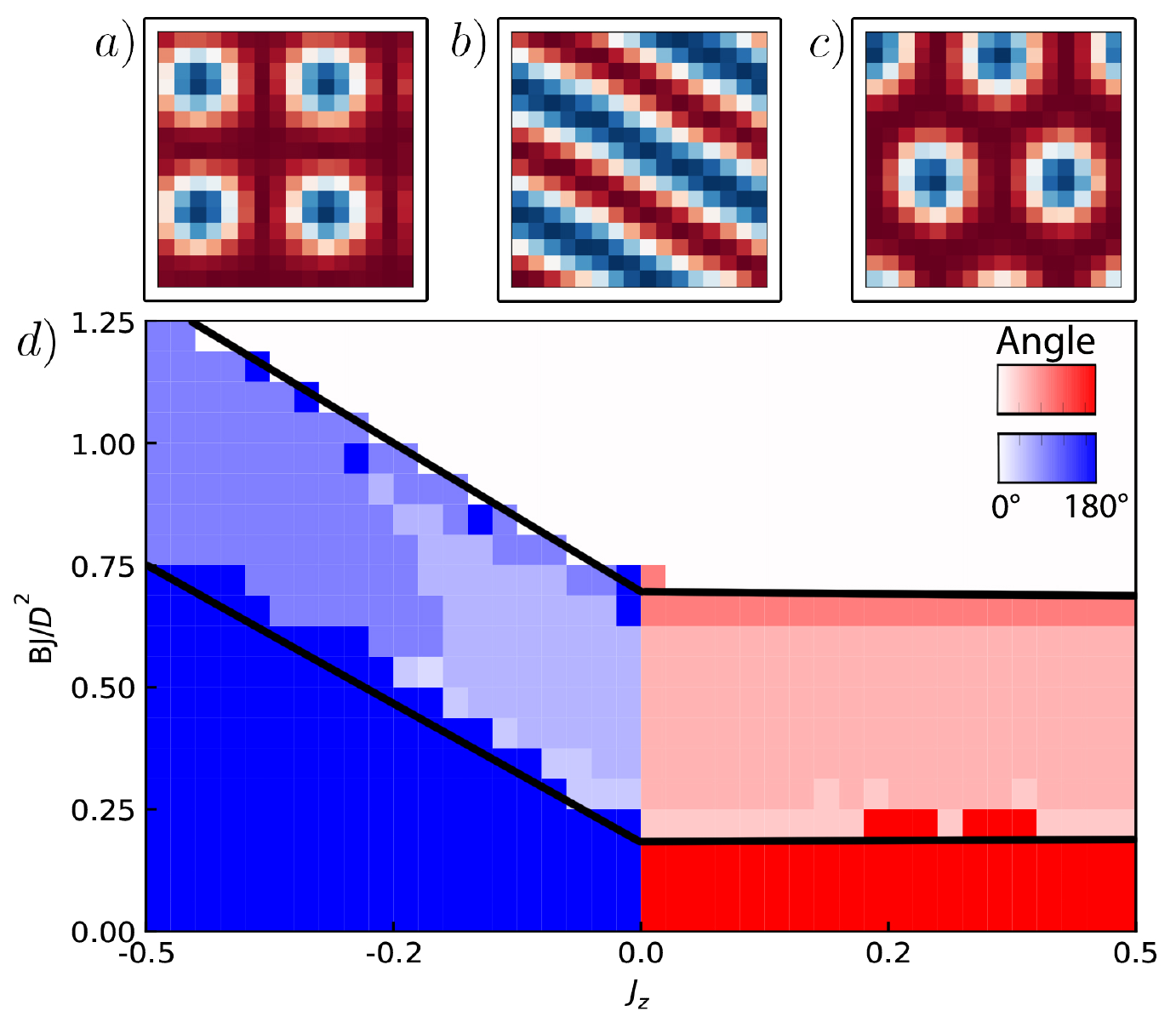}
\caption{Spin configurations of a 64x64 system showing a) square lattice configuration, b) spiral configuration and c) hexagonal lattice configuration. The colour in the snapshots depicts the z-component of the spin using the colorbar in Fig. 3. d) Phase diagram of a two layer SAF for changing interlayer exchange $J_{\rm z}/J_{\rm xy}$ and magnetic field $BJ_{\rm xy}/D^2$. The colour blue corresponds to antiferromagnetic ordering between the two layers, and red to ferromagnetic ordering. The boundary of the skyrmion phase is indicated by the black lines. The colour scale depicts the smallest angle found between two peaks in the 2D Fourier Transformation of the state. The striped phase b) gives an angle of 180$^\circ$, where the square lattice configuration a) gives 90$^\circ$ and the triangular lattice configuration c) 60$^\circ$. This phase diagram is calculated for a 16x16 system with a pitch of $p=8$ and $K=0$. }
\label{figConfigurations}
\end{figure}

In \cref{figConfigurations} the phase diagram for different intra-layer exchange and external magnetic field is shown. It is visible that the skyrmion pocket location is dependent on the magnetic field in the antiferromagnetic intra-layer exchange region as already shown in \cref{figInterlayerPhaseDiagram}. This leads to a larger spiral phase in the antiferromagnetic region. The hexagonal structural phase is dominant in the ferromagnetic part of the phase diagram and extends in the antiferromagnetic region: for a large antiferromagnetic intra-layer exchange the lattice configuration changes from hexagonal to square. This square skyrmion lattice occurs only in the antiferromagnetic part.

%

\section{Discussion  \label{secDisc}}
In this section, we turn to a physical interpretation of our results. For small skyrmions, skyrmions consisting of a small number of spins, the skyrmion lattice has a preferred orientation with respect to the underlying lattice, this can indicate that the skyrmion lattice is influenced by pinning to the underlying lattice. To test whether the square lattice configurations originate from this effect we increased the skyrmion pitch by lowering the DMI in our simulations. We found that the skyrmion lattices orient independently of the underlying lattice and that the skyrmion lattice still possesses the square configuration for increased skyrmion pitch.
Moreover, the periodic boundary conditions do not seem to influence the found results. Changing the aspect ratio of our systems or changing the periodic boundary conditions to twisted boundary conditions leads to no significant change in the skyrmion lattice configurations. 

\begin{figure}
\includegraphics[width=\linewidth]{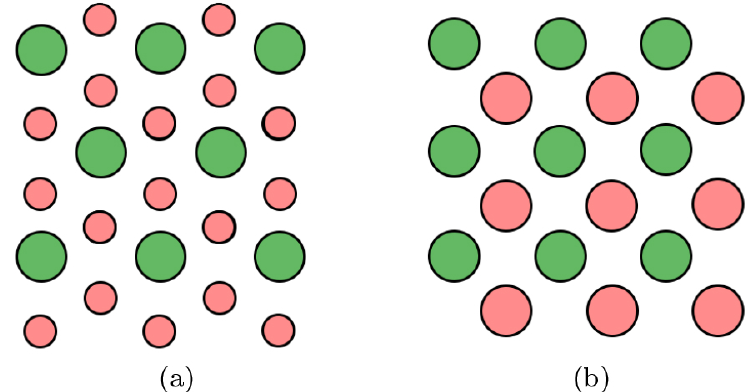} 
\caption{Schematic display of (a) a triangular lattice (green) together with its dual lattice (red) (b) a square lattice (green) with its dual lattice (red). The dual lattice of triangular lattice (a) has a higher density than the triangular lattice self, while the dual lattice of the square lattice (b) has an equal density.}
\label{figLattices}
\end{figure}

\begin{figure}
\includegraphics[width=\linewidth]{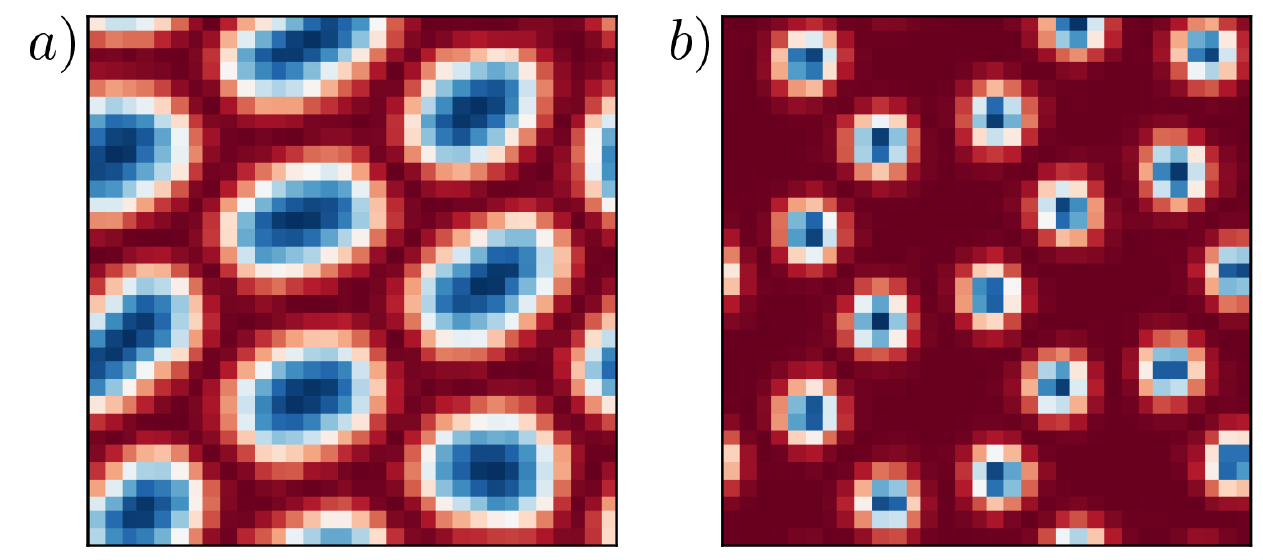} 
\caption{Snapshot of a spin configuration in both layers with $32$x$32$ spins, $J_{\rm z}= -0.20$, $B=0.65$, $K=0$, the colour depicts the $z$-component of the spin (see colorbar \cref{figFFT}). Layer a) has a pitch of $p=10$ and layer b) $p=6.25$. We see a triangular lattice in a) and a honeycomb lattice in b).}
\label{figHoney}
\end{figure}

The phase transition between the hexagonal and square skyrmion lattice can be explained as follows. The spin in the centre of the skyrmion points in the same direction in both layers, i.e. along the external magnetic field. Due to the antiferromagnetic interlayer exchange two skyrmions in different layers favour to be not on top of each other. Therefore, the skyrmion lattices are shifted relative to each other such that the skyrmions are positioned such that the spins in the different layers point oppositely. 
This implies that the skyrmion lattices in different layers are each other's dual lattice. The dual layer of a hexagonal lattice is a honeycomb lattice, and the dual lattice of square lattice is a square lattice again. It is not possible with equal skyrmion densities to form a honeycomb lattice in the dual lattice of the hexagonal lattice since the skyrmions will overlap. For a square lattice it is possible to position the skyrmions in their own dual lattice, which gives an energetic advantage over the triangular lattice. For a schematic display see \cref{figLattices}.
To test our interpretation we change the skyrmion density in one of the layers by altering the Dzyaloshinskii-Moriya interaction in that layer. By increasing the DMI, the skyrmion size decreases and the skyrmion density increases. Our interpretation suggests that the skyrmion lattice can have a honeycomb configuration in the layer with altered DMI because the smaller skyrmions do not overlap anymore. In \cref{figHoney} a snapshot is plotted of a spin configuration for parameters in the synthetic antiferromagnetic-square lattice regime but with different DMI for the two layers. Here the triangular lattice configuration manifests in the lower density layer and the honeycomb lattice in the higher density layer, confirming our interpretation of the structural phase transition.  

As seen in \cref{figConfigurations}, the structural phase transition occurs around $J_{\rm z}=-0.25$ for $p=8$. By looking at simulations with larger systems and skyrmion pitch sizes we noted that the phase transition occurs at different values of intra-layer exchange dependent on the pitch size of the skyrmions. For larger pitch size the phase transition is at higher intra-layer exchange, e.g. for skyrmion pitch $p=13$ the phase transition is around $J_{\rm z}=-0.02$. While our finding of the existence of the structural transition appears thus unchanged for larger system sizes, the precise value of $J_{\rm z}$ where the transition occurs is dependent on skyrmion size. Because of the excessive simulation time, we have not attempted finite-size scaling to determine the thermodynamic limit of the value of $J_{\rm z}$ where the transition occurs.


\section{Conclusion and Outlook \label{secConcl}}
In this article we studied skyrmions in synthetic antiferromagnets. We found that skyrmions still occur in bilayer synthetic antiferromagnets, but in the synthetic antiferromagnetic case a higher external magnetic field is needed than in the synthetic ferromagnetic case or the case without intralayer coupling. Skyrmion lattices in the synthetic antiferromagnetic case shift from a triangular to a square lattice for increasing interlayer coupling. Both the interlayer coupling and the external field have influence on the lattice geometry and both parameters are experimentally adjustable through either spacer thickness, external electric field, and through external magnetic field. 
As seen in \cref{figConfigurations}, the phase transition between the hexagonal synthetic antiferromagnetic part and the square synthetic antiferromagnetic part has a magnetic field dependence which gives an opportunity for applications. 

Different skyrmion lattices configurations such as the hexagonal-honeycomb configurations mentioned in  \cref{secDisc} could be explored further. For example, the phase diagram could be determined. 

Experimental verification of triangular-honeycomb configurations can be obtained by multilayer systems with different DMI in both layers. This can be engineered by creating two layers with different compositions \cite{Hrabec2014}. Direct observations of skyrmion lattices can be obtained by spin polarised scanning tunnelling microscopy or Lorentz transmission electron microscopy \cite{Bergmann2014,Yu2013}. 
The phase transition between the two lattice configurations is dependent on the ratio between the interlayer and intralayer exchange, and the ratio between the DMI and intralayer exchange.
For a pitch size of $p=13$ the interlayer- intralayer exchange ratio is $J_{\rm z}/J_{\rm xy} \sim 0.007$. This value is experimentally achievable \cite{Duine2017}.

This work is part of the research programme Skyrmionics - towards skyrmions for nanoelectronics, which is financed by the Netherlands Organisation for Scientific Research (NWO).

\bibliography{My_Collection}
\end{document}